# Duration of the Process of Complete Synchronization of Two Coupled Identical Chaotic Systems

A. A. Koronovskiĭ*, A. E. Hramov**, and I. A. Khromova***

*State Scientific Center "College", Saratov State University, Saratov, Russia*
*e-mail: * alkor@cas.ssu.runnet.ru; ** aeh@cas.ssu.runnet.ru; *** KhromovaIA@yandex.ru*

**Abstract**—We consider the time required for complete synchronization of two identical one-way coupled van der Pol–Duffing oscillators occurring in the regime of dynamic chaos. The influence of the initial phase difference between oscillators on the duration of the process of complete synchronization has been studied. At a fixed phase of chaotic oscillations of the self-excited drive oscillator, the period of time (past the coupling onset) during which the complete synchronization regime is established depends on the phase of the self-excited response oscillator.

The phenomenon of synchronization of self-excited oscillatory systems attracted the attention of researchers for a long time and is still an important problem in the modern theory of nonlinear oscillations and waves (see, e.g., [1–6]). In recent years, the duration of the synchronization process has been extensively studied for dynamical systems exhibiting both periodic and chaotic behavior [7–11]. The interest in this problem is related, in particular, to the extensively developing direction of research devoted to the use of chaotic signals for data transmission [12–14].

The time required for the complete (full) synchronization after the onset of external action can be considered as the duration of a transient process in a nonautonomous dynamical system. As is well known, the transient process duration in a dynamical system depends on the initial conditions and obeys certain laws [15–17]. In self-excited oscillators exhibiting periodic behavior, the initial conditions can be represented by the phase of oscillations. In particular, it was recently demonstrated [9] for self-excited oscillatory dynamical systems both with finite (van der Pol oscillator) and with infinite (distributed active medium of the "helical electron beam–backward electromagnetic wave" type) number of degrees of freedom under an external periodic action, that the duration of synchronization at the drive frequency significantly depends on the initial phase difference of oscillations in the response (slave) and drive (master) systems. Similar results were obtained for the time of complete synchronization of one-way coupled identical subsystems occurring in the regime of periodic oscillations [11].

In cases when a response dynamical system occurs in the regime of chaotic oscillations, the phase of these oscillations determined by one or another method [5, 18–20] does not uniquely determine the state of the system, while still being a convenient characteristic. This study was aimed at establishing how does the time of complete synchronization of two identical one-way coupled chaotic subsystems depends on the initial phase difference under otherwise variable initial conditions.

We have studied a model system comprising one-way coupled van der Pol–Duffing oscillators [21, 22]. As is known [23, 24], one-way coupled oscillators of this type (Fig. 1a) exhibit the phenomenon of complete chaotic synchronization. The drive oscillator is described by a system of dimensionless differential equations,

$$\dot{x} = -\nu[x^3 - \alpha x - y], \quad \dot{y} = x - y - z, \quad \dot{z} = \beta y, \quad (1)$$

and the response oscillator is described the system

$$\dot{u} = -\nu[(u)^3 - \alpha u - v] + \nu\varepsilon(x - u)\sigma(\tau - \tau_0), \\ \dot{v} = u - v - w, \quad \dot{w} = \beta v. \quad (2)$$

Here, $\sigma(\xi)$ is the Heaviside function and $(x, y, z)$ and $(u, v, w)$ are the dynamical variables characterizing the states of the drive and response oscillators, respectively. These quantities are related to the dimensional values by the following formulas [23, 24]: $x = V_1\sqrt{bR}$; $y = V_2\sqrt{bR}$; $z = i_L\sqrt{bR^3}$; $u = U_1\sqrt{bR}$; $v = U_2\sqrt{bR}$; $w = i'_L\sqrt{bR^3}$.

The current–voltage characteristic of the nonlinear element $N$ is described by a cubic parabola

$$i(V) = aV + bV^3. \quad (3)$$

The dimensionless time in the system of Eqs. (1) and (2) is defined as $\tau = t/(RC_2)$, and the dimensionless quantities $\alpha$, $\beta$, and $\nu$ are related to the dimensionless parameters of oscillators as $\alpha = -(1 + aR)$, $\beta = C_2 R^2/L$, and $\nu = C_2/C_1$; the coupling parameter is $\varepsilon = R/R_c$.

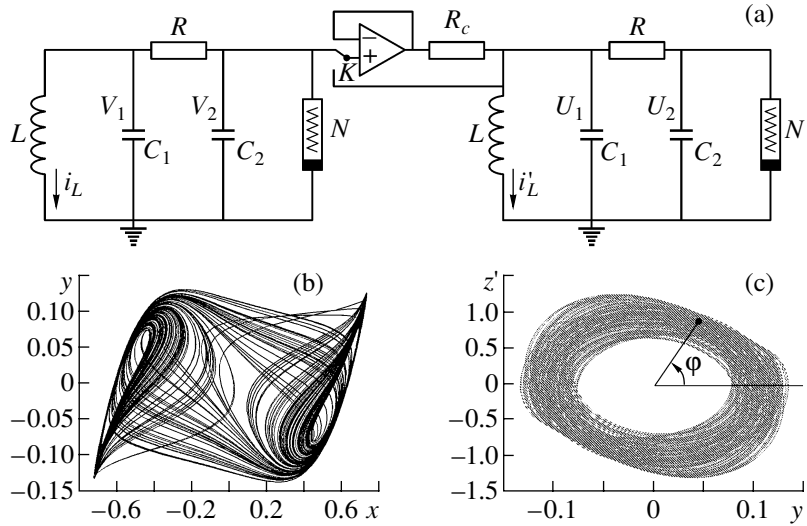

**Fig. 1.** A system of one-way coupled van der Pol–Duffing oscillators: (a) schematic diagram; (b) phase portrait of the drive oscillator (1) in a regime of bistable chaos with double-band chaotic attractor for the control parameters $\alpha = 0.35$, $\beta = 300$, and $\nu = 100$; (c) projection of the chaotic attractor onto the $(y, z')$ plane for determining the phase $\varphi$ of chaotic oscillations.

The system of equations describing one-way coupled oscillators (1) and (2) was numerically solved using the fourth-order Runge–Kutta method with an integration time step of $\Delta\tau = 2 \times 10^{-5}$. The control parameters were set (by analogy with [24]) equal to $\alpha = 0.35$, $\beta = 300$, and $\nu = 100$, for which the system exhibits bistable chaotic dynamics (Fig. 1b). The coupling parameter was selected equal to $\varepsilon = 1.35$, which ensures settling of the regime of complete synchronization of the two subsystems [24].

The one-way coupling of the self-excited drive and response oscillators was switched on at a time $\tau_0$ when the transient processes in both subsystems were accomplished and the imaging points in the phase spaces $(x, y, z)$ and $(u, v, w)$ reached the corresponding chaotic attractors. Prior to this moment, both oscillators occurred in the state of autonomous self-excited oscillations. The period of time $T_s$ required for the complete synchronization was determined as

$$T_s = \tau_s - \tau_0, \quad (4)$$

where $\tau_s$ is the time corresponding to the onset of complete synchronism. A criterion for the complete synchronization was selected in the following form:

$$\sqrt{(x-u)^2 + (y-v)^2 + (z-w)^2} \leq 5 \times 10^{-3}, \quad (5)$$
$$\forall \tau \geq \tau_s.$$

The phase $\varphi$ of chaotic oscillations for each oscillator was defined as the angle of projection of the chaotic attractor onto the plane of a new coordinate system: $(y' = y, z' = -0.8192x + 0.5735z)$ for the drive oscillator (Fig. 1c) and $(v' = v, w' = -0.8192u + 0.5735w)$ for the response oscillator (see [5, 18, 19]).

In the course of numerical simulation of Eqs. (1) and (2), the coupling between oscillators was always switched on at the same moment of time $\tau_0$, when the dynamical variables characterizing the state of the drive system (1) were $x = 0.041009$, $y = -0.09994$, $z = 0.75978$ that corresponded to the phase of chaotic oscillations in the drive (master) system $\varphi_m = 3\pi/4$. The initial conditions $(u, v, w)$ for the response oscillator (2) at the moment of coupling onset were taken different for each simulation, but so that the imaging point would belong to a chaotic attractor and the initial phase $\varphi_s$ of the response (slave) system would also be fixed.

The chaotic attractor of each system has two "bands ($x < 0$ and $x > 0$) corresponding to bistable chaotic dynamics, whereby the system exhibits switching between the two states (Fig. 1b). For this reason, the duration of synchronization for each initial slave phase $\varphi_s$ was studied separately for the upper and lower band of the chaotic attractor. The initial conditions for the drive oscillator were always taken the same (see above), corresponding to the upper (positive) band of the chaotic attractor.

The results of numerical simulations showed that the time of complete synchronization for the dynamical systems with bistable chaotic attractors depends on whether the initial conditions in the drive and response oscillators belong to the like or unlike bands of the corresponding chaotic attractor. When the initial conditions are selected on different bands (e.g., the upper band for the drive oscillator and the lower band for the response oscillator) of the chaotic attractor, the duration of complete synchronization depends to a significant extent on the phase of chaotic oscillations in the response subsystem (for a fixed initial phase of the drive oscillator).

These results are illustrated in Figs. 2a and 2b showing the distribution of synchronization times $T_s$ for different initial phases $\varphi_s$ of chaotic oscillations in the response oscillator. Each distribution is constructed for 100 sets of initial conditions (with the same initial phase) belonging to the lower band of the chaotic attractor. As can be seen, the shape of distributions is close to the $\delta$ function. This indicates that the time of complete synchronization in the system described by Eqs. (1) and (2) is determined by the phase of chaotic oscillations. For various initial conditions (belonging to the lower band of the chaotic attractor) corresponding to the same initial slave phase $\varphi_s$, the time of complete synchronization is also virtually the same.

A somewhat different pattern is observed when the initial conditions for the response oscillator at the coupling onset time $\tau_0$ are selected so that the imaging points fall within the like (in this case, the upper) bands of bistable chaotic attractors. If the initial phases of the drive ($\varphi_m$) and response ($\varphi_s$) oscillators differ rather significantly, the distribution of synchronization times, like that observed for the initial conditions belonging to the unlike bands, has a shape close to the $\delta$ function (Fig. 2c). However, should the initial phases of the drive and response oscillators be sufficiently close (for the imaging points in the like bands of the chaotic attractors), the distribution of synchronization times "scatters" (Fig. 2d). In this case, the time of complete synchronization of two one-way coupled subsystems varies within rather broad limits for the same initial phase of chaotic oscillations of the response subsystem.

This phenomenon is probably explained as follows. When the imaging points corresponding to the dynamics of drive and response oscillators have close initial phases and occur in the like bands of chaotic attractors, the subsystems can be sufficiently close to the state of complete synchronism. In this case, the time of complete synchronization will be much shorter than that for the same oscillators in the initial states significantly far from synchronism. However, the fact that the imaging points have close initial phases and belong to the like bands by no means implies that the two subsystems are necessarily close to synchronism, since it is known that chaotic dynamics is characterized by exponential expansion of the phase trajectories [25]. Therefore, the synchronization time $T_s$ will be comparable with that for the other initial slave phases $\varphi_s$. For this reason, the distribution f synchronization ties $T_s$ acquires the form depicted in Fig. 2d.

Thus, the time of synchronization of one-way coupled van der Pol–Duffing oscillators occurring in the regime of chaotic oscillations depends on the initial phase $\varphi_s$ of chaotic oscillations in the response oscillator tor (at a fixed initial phase $\varphi_m$ of the drive oscillator). Another important factor is whether the imaging points corresponding to the states of the drive and response oscillators occur in the like or unlike bands of the bistable chaotic attractors at the time $\tau_0$ of switching on the one-way coupling between the subsystems. When the imaging points belong to the unlike bands, the time $T_s$ of complete synchronization is virtually the same for all initial conditions corresponding to the same slave phase $\varphi_s$. However, should the imaging points of the two oscillators occur in the like bands of their chaotic attractors and have not strongly different initial phases at the moment of coupling onset, the average time of complete synchronization decreases and the shape of the $T_s$ distribution becomes different from the $\delta$ function characteristic of the situations with other initial slave phases $\varphi_s$.

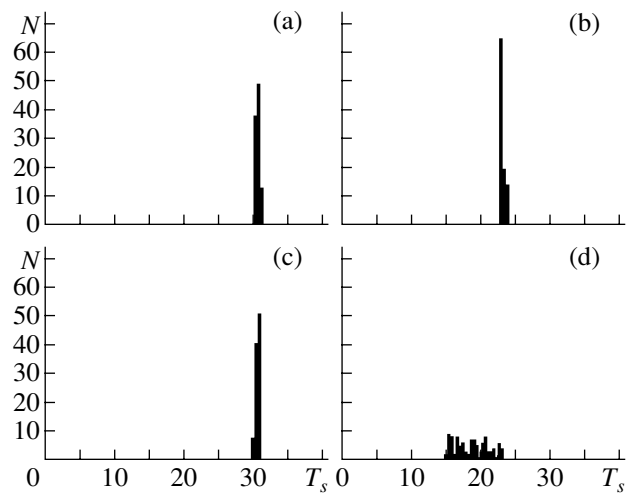

**Fig. 2.** Distributions of the time of complete synchronization of one-way coupled van der Pol–Duffing oscillators. The initial conditions for the drive oscillator are selected in the upper band of the bistable chaotic attractor and correspond to the initial phase $\varphi_m = 3\pi/4$. The initial conditions for the response oscillator are as follows: (a) lower band, $\varphi_s = 1.03$; (b) lower band, $\varphi_s = 1.73$; (c) upper band, $\varphi_s = 1.03$; (d) upper band, $\varphi_s = 2.78$. Each distribution is constructed for 100 sets of initial conditions randomly selected on the corresponding chaotic attractor.

**Acknowledgments.** This study was supported by the Russian Foundation for Basic Research (project no. 02-02-16351) and jointly by the Scientific-Education Center "Nonlinear Dynamics and Biophysics" at the Saratov State University and the U.S. Civilian Research and Development Foundation for the Independent States of the Former Soviet Union (CRDF grant no. REC-006).